\documentclass[prl,aps,showpacs,onecolumn,floatfix]{revtex4}
\usepackage{epsfig} \usepackage{graphics} \usepackage{bm}
\usepackage{amssymb}
\usepackage{graphicx}
\begin{document}

\preprint{Lebed-IJMPD}

\title{BREAKDOWN OF THE EQUIVALENCE BETWEEN GRAVITATIONAL MASS AND
ENERGY FOR A QUANTUM BODY: THEORY AND SUGGESTED EXPERIMENTS}

\author{Andrei G. Lebed$^*$}

\affiliation{Department of Physics, University of Arizona, 1118 E.
4-th Street, Tucson, Arizona 85721, USA; \
lebed@physics.arizona.edu}

\begin{abstract}
We review recent theoretical results, obtained for the equivalence
between gravitational mass and energy of a composite quantum body
as well as for its breakdown at macroscopic and microscopic
levels. In particular, we discuss that the expectation values of
passive and active gravitational masses operators are equivalent
to the expectation value of energy for electron stationary quantum
states in a hydrogen atom. On the other hand, for superpositions
of the stationary quantum states, inequivalence between the
gravitational masses and energy appears at a macroscopic level. It
reveals itself as time-dependent oscillations of the expectation
values of passive and active gravitational masses, which can be,
in principle, experimentally measured. Inequivalence between
passive gravitational mass and energy at a microscopic level can
be experimentally observed as unusual electromagnetic radiation,
emitted by a macroscopic ensemble of the atoms. We propose the
corresponding experiment, which can be done on the Earth's orbit,
using small spacecraft. If such experiment is done it would be the
first direct observation of quantum effects in general relativity.
\end{abstract}

\pacs{04.60.-m, 04.80.Cc}

\maketitle


\section{1. Introduction}
Unification theory of all interactions in nature has been
considered as one of the major problems in physics for many years.
The idea to unify all existing forces is behind such successful
theories as the Maxwell's electromagnetism, electroweak theory of
Salam, Weinberg, and Glashow and finally - the so-called standard
theory of elementary particles. What is obviously missing here is
gravitational force, which has not been unified with quantum
mechanics so far. Note that the current situation with possible
unification of gravitation theory and quantum mechanics is very
unclear. This is partially due to the fact that the fundamentals
of these two theories are very different from each other and
partially due to the lack of the corresponding experimental data.
Normally, one can expect that quantum gravity reveals itself in
full at the so-called Planck energy of the order of $10^{28}eV$,
which is much higher than the energy range accessible to
experimentalists. On the other hand, if one is interested in
quantum effects due to quantization of matter only, they can
appear at accessible energies (see, for example, neutron
experiments [1,2]). Note that, despite of their significant
importance, the experiments [1,2] have an obvious drawback: they
study quantum mechanical effects in Newtonian gravity, where all
general relativity effects are negligible. Below, we review two
recently suggested quantum phenomena in general relativity [3-6],
which have to be accessible at the existing energies, as well as
two corresponding idealized experiments to discover these
phenomena. It is also important that the above mentioned quantum
effects in general relativity significantly break the textbook
equivalence between gravitational mass of a composite quantum body
and its energy, which also can be experimentally tested.

It is known that the notion of gravitational mass of a composite
classical body is not a trivial one. Moreover, it is related to
the following interesting paradoxes. Indeed, if we apply the
so-called Tolman's formula for active gravitational mass [7],
\begin{equation}
m_a^g = \frac{1}{c^2} \int [T_0^0({\bf r}) - T_1^1({\bf r}) -
T_2^2({\bf r}) -T_3^3({\bf r})] \ d^3 r \ ,
\end{equation}
to a free photon with energy $E$, we obtain $m^g_a = 2E/c^2$
(i.e., two times bigger value than the expected one)[8]. Let us
now consider the photon in a box with mirror walls (i.e., a
composite body at rest). Then, as shown by Misner and Putnam [8],
the Einstein's equation, $m^g_a=E/c^2$, restores, if we take into
account a negative contribution to active gravitational mass from
stress in the box walls. So, in the example above, both kinetic
and potential energies make contributions to gravitational mass
and the Einstein's equation (what is important) is restored only
after averaging over time. Nordtvedt [9] and Carlip [10]
considered more realistic example - classical model of a hydrogen
atom (i.e., positive and negative charges bound by the Coulomb
interaction). They showed that kinetic energy, $K$, and potential
energy, $U$, are coupled to external gravitational field in
different ways. More specifically, they demonstrated [9,10] that
the following combination, $3K + 2U$, is coupled with the
gravitational field, which is rather unexpected. Nevertheless, the
Einstein's equation can be restored for this classical model of a
hydrogen atom after averaging over time [9,10]. Indeed, due to the
classical virial theorem, the following time averaging is zero:
\begin{equation}
<2K+U>_t=0.
\end{equation}
Therefore, if we consider averaged over time active and passive
gravitational masses of classical model of a hydrogen atom, we
obtain:
\begin{equation}
<m^g_{a,p}>_t = m_e + m_p + \frac{<3K + 2U>_t}{c^2}=\frac{E}{c^2},
\end{equation}
where $m_e$ and $m_p$ are the bare electron and proton masses,
respectively.

\section{2. Goal}

The goal of our review is to describe in detail the recent results
[3-6] related to passive and active gravitational masses of a
composite quantum body. As the simplest example of such a body, we
consider a hydrogen atom in external gravitational field. In
section 3, we reproduce the results of Refs.[9,10], related to
passive gravitational mass of classical model of a hydrogen atom,
using different method. Section 4 of the review is devoted to the
following quantum results: the expectation values of passive and
active gravitational masses are equivalent to the expectation
value of energy for stationary quantum states [3-6]. In section 5,
we review the results that demonstrate that the above mentioned
equivalence is broken for the corresponding expectation values for
superpositions of stationary quantum states [3,5]. This breakdown
happens at macroscopic level and reveals itself as time dependent
oscillations of the expectation values of passive and active
gravitational masses. In section 6, we review suggested in Ref.[5]
an idealized experiment to discover these oscillations. In
sections 7 and 8, we discuss breakdown of the equivalence between
passive gravitational mass and energy at a microscopic level for
stationary quantum states and the suggested idealized experiment
to discover this effect [3-6]. In section 9, we summarize our
results and discuss some still unresolved problems. Note that, if
one of the suggested in Refs. [3-6] and described in detail in the
review experiments is done, it would be the first direct
experiment to detect quantum effects in general relativity.

\section{3. Gravitational mass in classical physics}

In this section, we derive the Lagrangian and Hamiltonian for
classical model of a hydrogen atom in the Earth's gravitational
field [9,3]. Below, we account for terms only of the order of
$1/c^2$ and, thus, take into account only couplings of
non-relativistic kinetic and the Coulomb potential energies with
weak gravitational field. To be more specific, we disregard
electromagnetic and gravitational waves, magnetic force,
spin-orbital interaction, spin-spin interaction, etc. It is very
important that we disregard also all tidal effects. To derive the
corresponding Lagrangian, let us express the interval in the
Earth's gravitational field by means of the so-called weak field
approximation [11]:
\begin{equation}
d s^2 = -\biggl(1 + 2 \frac{\phi}{c^2} \biggl)(cdt)^2 + \biggl(1 -
2 \frac{\phi}{c^2} \biggl) (dx^2 +dy^2+dz^2 ), \ \phi = -
\frac{GM}{R} ,
\end{equation}
where $c$ is the velocity of light, $G$ is the gravitational
constant, $M$ is mass of the Earth, $R$ is distance between center
of the Earth and center of mass of a hydrogen atom (in our
approximation - proton). Note that, to calculate the Lagrangian
and Hamiltonian in a linear with respect to small parameter,
$|\phi(R)|/c^2 \ll 1$, approximation, we don't need to keep in
Eq.(4) the terms of the order $[\phi(R)/c^2]^2$, unlike the
classical problem about perihelion of the Mercury calculations
[11].

If we disregard all tidal effect, we can introduce the local
proper spacetime coordinates,
\begin{equation}
x'=\biggl(1-\frac{\phi}{c^2} \biggl) x, \ y'=
\biggl(1-\frac{\phi}{c^2} \biggl) y, \
z'=\biggl(1-\frac{\phi}{c^2} \biggl) z , \ t'=
\biggl(1+\frac{\phi}{c^2} \biggl) t,
\end{equation}
and write the Lagrangian and action for classical model of a
hydrogen atom in the following standard forms:
\begin{equation}
L' = -m_p c^2 -m_e c^2 + \frac{1}{2} m_e ({\bf v'})^2 +
\frac{e^2}{r'} \ , \ \ \ S' = \int L' dt' ,
\end{equation}
where $e$ is the electron charge, $r'$ is distance between
electron and proton, and ${\bf v'}$ is electron velocity. [Note
that, in this review, we use inequality $m_p \gg m_e$. Therefore,
we disregard kinetic energy of proton and consider its position as
center of mass of a hydrogen atom.] It is easy to show that the
Lagrangian (6) can be expressed in inertial coordinates
$(x,y,z,t)$ as
\begin{equation}
L = -m_p c^2 -m_e c^2 +  \frac{1}{2}m_e{\bf v}^2+\frac{e^2}{r} -
m_e \phi - \biggl( 3m_e\frac{{\bf v}^2}{2}-2\frac{e^2}{r} \biggl)
\frac{\phi}{c^2} .
\end{equation}

To calculate the Hamiltonian of classic model of a hydrogen atom
in the Earth's gravitational potential, we use the standard
procedure: $H({\bf p},{\bf r}) = {\bf p}{\bf v} - L({\bf v},{\bf
r})$, where ${\bf p} = \partial L({\bf v},{\bf r})/ \partial {\bf
v}$. After simple calculations, we obtain:
\begin{equation}
H = m_p c^2 + m_e c^2 + \frac{{\bf p}^2}{2m_e}-\frac{e^2}{r} + m_p
\phi + m_e \phi + \biggl( 3 \frac{{\bf p}^2}{2 m_e}
-2\frac{e^2}{r} \biggl) \frac{\phi}{c^2},
\end{equation}
where the canonical momentum in gravitational field is ${\bf
p}=m_e {\bf v} (1-3 \phi/c^2)$. [It is important that, in the
review, we disregard all tidal effects. This means that we
consider a hydrogen atom as a point-like body and do not
differentiate gravitational potential with respect to the
coordinates ${\bf r}$ and ${\bf r'}$, corresponding to electron
position in the center of mass coordinate system. It is possible
to show that in such a way we disregard tidal effects of the order
of $|\phi/c^2|(r_B/R) \sim 10^{-26}$, where $r_B$ is the so-called
Bohr radius (i.e., the typical size of a hydrogen atom).] Let us
introduce averaged over time electron passive gravitational mass,
$<m^g_p>_t$. From Eq.(8), it follows that
\begin{equation}
<m^g_p>_t = m_e  + \biggl<  \frac{{\bf p}^2}{2 m_e}-
\frac{e^2}{r}\biggl>_t \frac{1}{c^2} +\biggl< 2 \frac{{\bf p}^2}{2
m_e}-\frac{e^2}{r}\biggl>_t \frac{1}{c^2} =  m_e + \frac{E}{c^2} \
,
\end{equation}
where $E= {\bf p}^2/2 m_e - e^2/r$ is electron energy. We pay
attention to the fact that the third term in Eq.(9) is equal to
zero due to the classical virial theorem. Therefore, we reproduce,
using Hamiltonian approach, the results of Refs.[9,10]. Indeed, as
directly seen from Eq.(9) the averaged over time passive
gravitational mass of a composite body in classical physics is
related to its energy by the Einstein's equation.

\section{4. Equivalence of the expectation values of gravitational mass
and energy for stationary quantum states} In this section, we
review some of the general results of Refs.[3-6].

\subsection{4.1. Operator of passive gravitational mass}

Here, we start to consider a quantum problem of gravitational mass
of a composite body [3-6]. Let us quantize the Hamiltonian (8) by
 substituting momentum operator, $\hat {\bf p}=- \hbar \partial / \partial {\bf r}$, instead of canonical
 momentum, {\bf p}. For convenience, we represent the quantized
Hamiltonian in the following form:
\begin{equation}
\hat H = m_p c^2 + m_e c^2 + \frac{\hat {\bf
p}^2}{2m_e}-\frac{e^2}{r} + m_p \phi + \hat m^g_p \phi \ ,
\end{equation}
where we define electron passive gravitational mass operator as
operator, which is proportional to electron weight operator in the
weak gravitational field [3-6],
\begin{equation}
\hat m^g_p  = m_e + \biggl(\frac{\hat {\bf p}^2}{2m_e}
-\frac{e^2}{r}\biggl)\frac{1}{c^2} + \biggl(2 \frac{\hat {\bf
p}^2}{2m_e}-\frac{e^2}{r} \biggl) \frac{1}{c^2} \ .
\end{equation}
We pay attention to the fact that the first term in Eq.(11) is the
bare electron mass, the second term is the expected kinetic and
potential energies contributions to electron mass, whereas the
third term is the so-called virial one. It is important that the
virial operator in Eq.(11) does not commute with electron energy
operator, taken in the absence of gravitational field. Note that
Eq.(11) can be obtained directly from the Dirac equation in curved
spacetime in a weak gravitational field. Indeed, it is possible to
show that Eq.(11) can be derived from the Hamiltonian (3.24) of
Ref.[12], if we omit all tidal terms. Note that, in Ref.[12], a
completely different physical problem is considered.

\subsection{4.2. Equivalence of the expectation values of passive
gravitational mass and energy for stationary quantum states}

\subsubsection{4.2.1. Non-relativistic case}

Let us discuss an important consequence of Eq.(11) for electron
passive gravitational mass operator. Suppose that we have a
macroscopic ensemble of hydrogen atoms each of them being in a
ground state with energy $E_1$,
\begin{equation}
\Psi_1(r,t) = \Psi_1(r) \exp(-iE_1t/\hbar).
\end{equation}
Then, as directly seen from Eq.(11), the expectation value of the
electron gravitational mass operator is
\begin{equation}
<\hat m^g_p> = m_e + \frac{ E_1}{c^2}  + \biggl< 2 \frac{\hat {\bf
p}^2}{2m_e}-\frac{e^2}{r} \biggl> \frac{1}{c^2} = m_e +
\frac{E_1}{c^2}  ,
\end{equation}
where the third term in Eq.(13) is zero due to the quantum virial
theorem [13]. It is trivial to extend the obtained result (13) to
any stationary quantum state in a hydrogen atom. Thus, we can
conclude that the expectation values of passive gravitational mass
and energy are equivalent to each other for stationary quantum
states [3-6].

\subsubsection{4.2.2 Relativistic corrections}

In this subsection, we study a more general case [5], where we
include the so-called relativistic corrections to electron motion
in a hydrogen atom. If we disregard all tidal effects, we can
write the Schr\"{o}dinger equations in external gravitational
field in proper spacetime coordinates in the following way:
\begin{equation}
i \hbar \frac{\partial \Psi({\bf r'},t')}{\partial t'} = \hat H
(\hat {\bf p'},{\bf r'}) \Psi({\bf r'},t')  .
\end{equation}
It is important that the Hamiltonian (14) now contains both
non-relativistic and relativistic parts and can be written as [14]
\begin{equation}
\hat H (\hat {\bf p'},{\bf r'}) = \hat H_0 ( \hat {\bf p'},{\bf
r'}) + \hat H_1 (\hat {\bf p'},{\bf r'}) ,
\end{equation}
where
\begin{equation}
 \hat H_0 (\hat {\bf p'},{\bf r'}) = m_e c^2 + \frac{\hat {\bf p'}^2}{2m_e}
 -\frac{e^2}{r'}
\end{equation}
and
\begin{equation}
\hat H_1 (\hat {\bf p'},{\bf r'}) = \alpha \hat {\bf p'}^4 + \beta
\delta^3 ({\bf r'}) + \gamma \frac{\hat {\bf S} \cdot \hat {\bf
L'} }{(r')^3} ,
\end{equation}
with the following values of the parameters:
\begin{equation}
\alpha = - \frac{1}{8 m_e^3 c^2}, \ \beta=\frac{\pi e^2
\hbar^2}{2m_e^2c^2}, \ \gamma = \frac{e^2}{2 m_e^2 c^2}.
\end{equation}
[Here, $\delta^3({\bf r})$ is the three-dimensional Dirac's
delta-function, $\hat {\bf L'} = - i \hbar [{\bf r'} \times
\partial / \partial {\bf r'}]$ is electron angular momentum
operator.] We point out that the first term in Eq.(17) is a
relativistic correction to electron kinetic energy, the second
term (with delta function) is the so-called Darwin's term, and the
third one is the spin-orbital interaction. Using the coordinate
transformation (5), we can rewrite the Hamiltonian (15)-(18) in
inertial coordinate system $(x,y,z,y)$ in the following way:
\begin{equation}
\hat H(\hat {\bf p},{\bf r}, \phi)= [\hat H_0(\hat {\bf p}, {\bf
r}) + \hat H_1 (\hat {\bf p},{\bf r})] \biggl(1 + \frac{\phi}{c^2}
\biggl) + \biggl(2 \frac{\hat {\bf p}^2}{2 m_e}-\frac{e^2}{r} + 4
\alpha \hat {\bf p}^4 + 3 \beta \delta^3({\bf r}) + 3 \gamma
\frac{\hat {\bf S} \cdot \hat {\bf L} }{r^3} \biggl)
\frac{\phi}{c^2} .
\end{equation}
From Eq.(19) it follows that passive gravitational electron mass
operator with relativistic corrections can be written in a more
complicated than Eq.(11) form:
\begin{equation}
\hat m^g_p = m_e + \biggl( \frac{\hat {\bf p}^2}{2m_e} -
\frac{e^2}{r} + \alpha \hat {\bf p}^4 + \beta \delta^3 ({\bf r}) +
\gamma \frac{\hat {\bf S} \cdot \hat {\bf L} }{r^3} \biggl)/c^2 +
\biggl(2 \frac{\hat {\bf p}^2}{2 m_e}-\frac{e^2}{r} + 4 \alpha
\hat {\bf p}^4 + 3 \beta \delta^3 ({\bf r}) + 3 \gamma \frac{\hat
{\bf S} \cdot \hat {\bf L} }{r^3} \biggl)/c^2 .
\end{equation}
Suppose that we consider again a macroscopic ensemble of hydrogen
atoms with each of them being in ground quantum state with energy
$E'_1$,
\begin{equation}
\Psi_1(r,t) = \Psi_1(r) \exp(-iE'_1t/\hbar),
\end{equation}
where $E'_1$ is calculated in the presence of relativistic
corrections (17) to electron energy in a hydrogen atom. In this
case, the expectation value of electron passive gravitational mass
can be expressed as
\begin{equation}
<\hat m^g_p > = m_e + \frac{ E'_1}{c^2} + \biggl<2 \frac{\hat {\bf
p}^2}{2 m_e}-\frac{e^2}{r} + 4 \alpha \hat {\bf p}^4 + 3 \beta
\delta^3 ({\bf r}) + 3 \gamma \frac{\hat {\bf S} \cdot \hat {\bf
L} }{r^3} \biggl>/c^2 .
\end{equation}

Below, we demonstrate that the expectation value of the third term
in Eq.(22) iz zero and, thus, the equivalence between passive
gravitational mass and energy survives for stationary quantum
states in the presence of relativistic corrections [5]. To this
end, we define the so-called virial operator [13],
\begin{equation}
\hat G = \frac{1}{2} (\hat {\bf p}{\bf r} +{\bf r} \hat {\bf p}) ,
\end{equation}
and write a standard equation for motion of its expectation value,
\begin{equation}
\frac{d}{dt} \biggl< \hat G \biggl> = \frac{i}{\hbar} \biggl<
[\hat H_0(\hat {\bf p},{\bf r}) + H_1(\hat {\bf p},{\bf r}), \hat
G] \biggl> ,
\end{equation}
where $[\hat A, \hat B]$ is, as usual, a commutator of two
operators, $\hat A$ and $\hat B$. Note that, for the stationary
quantum state (21), the derivative $d <\hat G>/dt =0$ and,
therefore,
\begin{equation}
\biggl< [\hat H_0(\hat {\bf p},{\bf r})+ H_1(\hat {\bf p},{\bf
r}), \hat G] \biggl> = 0 ,
\end{equation}
where the Hamiltonians in Eqs.(24),(25) are defined by Eqs.
(16)-(18). Using straightforward but lengthly calculations, it is
possible to show that
\begin{equation}
\frac{[\hat H_0(\hat {\bf p},{\bf r}), \hat G]}{-i \hbar}= 2
\frac{\hat {\bf p}^2}{2m_e}-\frac{e^2}{r} , \ \ \frac{[\alpha \hat
{\bf p}^4,\hat G]}{-i \hbar} = 4 \alpha \hat {\bf p}^4 , \ \
\frac{[\beta \delta^3 ({\bf r}), \hat G]}{-i \hbar} = 3 \beta
\delta^3 ({\bf r}), \ \ \frac{1}{-i \hbar} \biggl[ \gamma
\frac{\hat {\bf S} \cdot \hat {\bf L}}{r^3} , \hat G \biggl] = 3
\gamma \frac{\hat {\bf S} \cdot \hat {\bf L}}{r^3} ,
\end{equation}
where we use the following property of the Dirac's delta function:
$x[d \delta(x)/dx] = - \delta(x)$. As directly seen from
Eqs.(25),(26),
\begin{equation}
\biggl<2 \frac{\hat {\bf p}^2}{2 m_e}-\frac{e^2}{r} + 4 \alpha
\hat {\bf p}^4 + 3 \beta \delta^3 ({\bf r}) + 3 \gamma \frac{\hat
{\bf S} \cdot \hat {\bf L} }{r^3} \biggl> =0 ,
\end{equation}
and, thus, Eq.(22) satisfies the Einstein's equation:
\begin{equation}
<\hat m^g_p > = m_e + \frac{E'_n}{c^2} .
\end{equation}
It is important that Eq.(27) extends the so-called relativistic
quantum virial theorem [15] to the case, where the particles have
spin $1/2$.

Note that in this subsection we have established the equivalence
between the expectation values of passive gravitational mass of a
hydrogen atom and its energy in the presence of the relativistic
corrections to electron energy [5]. Here, we put forward a
hypothesis that such equivalence between the expectations values
of passive gravitational mass of a quantum system and its energy
always exists for stationary quantum states.

\subsection{4.3. Equivalence of the expectation values of active gravitational
mass and energy for stationary quantum states} In this subsection,
we review some general results, obtained in Ref.[5].

\subsubsection{4.3.1. Active gravitational mass in classical physics}
Here, we introduce active gravitational mass for classical model
of a hydrogen atom, where we have light negatively charged
particle (i.e., electron) exhibiting a bound motion in the Coulomb
electrostatic field of heavy positively charge particle (i.e.,
proton). At large distances from the atom, $R \gg r_B$,
gravitational potential in the first approximation is
 \begin{equation}
\phi(R)=-G \frac{m_p + m_e}{R} \ ,
\end{equation}
where so far we have not taken into account electron kinetic and
potential energies contributions. Since $m_p \gg m_e$, as usual,
we disregard kinetic energy of proton and consider its position as
center of mass of the atom. Now, we are in a position to calculate
how kinetic and potential energies of electron contribute to
electron active gravitational mass. To this end, we define active
gravitational mass of the atom by evaluating gravitational
potential, which act on a small test body located at distance much
high than the effective "size" of the atom, $r_B$. For sake of
simplicity we prescribe the Coulomb potential energy to electron
and, thus, consider the corresponding corrections to electron
active gravitational mass. As it follows from general theory of a
weak gravitational field [7,11], gravitational potential for our
case can be written as
\begin{equation}
\phi(R,t)=-G \frac{m_p + m_e}{R}- G \int \frac{\Delta
T^{kin}_{\alpha \alpha}(t,{\bf r})+ \Delta T^{pot}_{\alpha
\alpha}(t,{\bf r})}{c^2R} d^3 {\bf r} ,
\end{equation}
where $\Delta T^{kin}_{\alpha \beta}(t,{\bf r})$ and $\Delta
T^{pot}_{\alpha \beta}(t,{\bf r})$ are changes of stress-energy
tensor component $T_{\alpha \beta}(t, {\bf r})$ due to kinetic and
the Coulomb potential energies, correspondingly. We note that in
Eq.(30) only terms of the order of $1/c^2$ are taken into account,
which, for example, means that we disregard all retardation
effects. Therefore, in this approximation, electron active
gravitational mass is equal to
\begin{equation}
m^g_a = m_e + \frac{1}{c^2} \int [\Delta T^{kin}_{\alpha
\alpha}(t,{\bf r}) + \Delta T^{pot}_{\alpha \alpha}(t,{\bf r})]
d^3{\bf r}.
\end{equation}

To calculate $\Delta T^{kin}_{\alpha \alpha}(t, {\bf r})$, we use
the standard expression for stress-energy tensor of a moving
relativistic point mass [7,11]:
\begin{equation}
T^{\alpha \beta}({\bf r},t) = \frac{m v^{\alpha}(t)
v^{\beta}(t)}{\sqrt{1-v^2(t)/c^2}} \ \delta^3[{\bf r}-{\bf
r}_e(t)],
\end{equation}
where $v^{\alpha}$ is a four velocity and ${\bf r}_e$ is electron
trajectory in three dimensional space. From Eq.(32), it directly
follows that for low enough velocities, $v \ll c$,
\begin{equation}
\Delta T^{kin}_{\alpha \alpha}(t) = \int \Delta T^{kin}_{\alpha
\alpha}(t,{\bf r}) d^3{\bf r} = 3 \frac{mv^2(t)}{2}.
\end{equation}
Now, let us write the standard formula for stress energy tensor of
electromagnetic field,
\begin{equation}
T_{em}^{\mu \nu} = \frac{1}{4 \pi} [F^{\mu \alpha} F^{\nu}_{\
\alpha} - \frac{1}{4} \eta^{\mu \nu} F_{\alpha \beta} F^{\alpha
\beta}],
\end{equation}
where $\eta_{\alpha \beta}$ is the Minkowski metric in the absence
of gravitational field, $F^{\alpha \beta}$ is the so-called tensor
of electromagnetic field [7]. We stress that, in this subsection,
we take into account only the Coulomb electric field without
relativistic corrections. In this case, Eq.(34) can be
significantly simplified and, as a result, we obtain the following
expression for change of the electromagnetic stress energy tensor:
\begin{equation}
\Delta T^{pot}_{\alpha \alpha} (t) = \int \Delta T^{pot}_{\alpha
\alpha}(t,{\bf r}) d^3{\bf r} = -2\frac{e^2}{r(t)}.
\end{equation}
From Eqs.(33),(35), it follows that active electron gravitational
mass can be written in a similar way as the passive one:
\begin{equation}
m^g_a = m_e + \biggl(\frac{m_e{\bf
v}^2}{2}-\frac{e^2}{r}\biggl)/c^2 + \biggl(2\frac{m_e {\bf
v}^2}{2}-\frac{e^2}{r}\biggl)/c^2.
\end{equation}
Note that the last term in Eq.(36) is the virial one, which
depends on time. Therefore, we can conclude that active
gravitational mass of a classical model of a hydrogen atom depends
on time too. Nevertheless, it is possible to restore the
Einstein's equation for the equivalence between the mass and
energy. To this end, we introduce active gravitational mass of
electron, averaged over time [9,10]:
\begin{equation}
<m^g_a>_t = m_e + \biggl< \frac{m_e {\bf v}^2}{2}-\frac{e^2}{r}
\biggl>_t /c^2 + \biggl< 2 \frac{m_e {\bf
v}^2}{2}-\frac{e^2}{r}\biggl>_t /c^2 =m_e + E/c^2,
\end{equation}
where the averaged over time virial term is zero due to the
classical virial theorem.

\subsubsection{4.3.2. Active gravitational mass in quantum physics}

Let us first rewrite Eq.(36) for active electron gravitational
mass using momentum, istead of velocity,
\begin{equation}
m^g_a = m_e +\biggl(\frac{{\bf
p}^2}{2m_e}-\frac{e^2}{r}\biggl)/c^2 + \biggl(2\frac{{\bf
p}^2}{2m_e}-\frac{e^2}{r}\biggl)/c^2,
\end{equation}
and let us quantize the expression (38):
\begin{equation}
\hat m^g_a = m_e +\biggl(\frac{{\bf \hat
p}^2}{2m_e}-\frac{e^2}{r}\biggl)/c^2 + \biggl(2\frac{{\bf \hat
p}^2}{2m_e}-\frac{e^2}{r}\biggl)/c^2.
\end{equation}
In this subsection, we make use of semi-classical approach to
quantum general relativity [16], where gravitational field is not
quantized but matter is quantized,
\begin{equation}
R_{\mu \nu} - \frac{1}{2}R g_{\mu \nu} = \frac{8 \pi G}{c^4}
\bigl<\hat T_{\mu \nu} \bigl> ,
\end{equation}
where $<\hat T_{\mu \nu}>$ stands for the expectation value of
quantum operator, corresponding to the stress energy tensor. From
Eqs.(39),(40), it follows that the expectation value of electron
active gravitational mass is
\begin{equation}
<\hat m^g_a> = m_e +\biggl< \frac{{\bf \hat
p}^2}{2m_e}-\frac{e^2}{r}\biggl>/c^2 + \biggl<2\frac{{\bf \hat
p}^2}{2m_e}-\frac{e^2}{r}\biggl>/c^2,
\end{equation}
where, as usual, the last term is the virial one. Now, let us
suppose that we have a macroscopic ensemble of hydrogen atoms with
each of them being in ground state (12). Then, the expectation
value of electron active gravitational mass operator is equal to
\begin{equation}
<\hat m^g_a> = m_e + \frac{E_1}{c^2},
\end{equation}
where we take into account that the expectation value of the
virial term in Eq.(41) is equal to zero in stationary quantum
states due to the quantum virial theorem [13]. Therefore, we can
make the statement that, in stationary quantum states, active
gravitational mass of a composite quantum body is equivalent to
its energy [5].

\section{5. Inequivalence between gravitational mass and energy for
superposition of stationary quantum states} In this section, we
review the results of Refs.[3,5] that, in superpositions of
stationary quantum states, where the expectation values of energy
do not depend on time, the expectation values of passive and
active gravitational masses exhibit time dependent quantum
oscillations. These oscillations represent novel quantum
phenomenon. Moreover, their observation would be the first
observation of quantum effects in general relativity. In section
6, we suggest the corresponding idealized experiment.

\subsection{5.1. Inequivalence between passive gravitational mass and
energy} Let us consider the simplest superposition of stationary
quantum states in a hydrogen atom - the superposition of the
ground and first s-wave excited states (i.e., 1S and 2S electron
orbits),
\begin{equation}
\Psi_{1,2}(r,t) = \frac{1}{\sqrt{2}} \bigl[ \Psi_1(r) \exp(-iE_1t)
+ \Psi_2(r) \exp(-iE_2t) \bigl].
\end{equation}
It is easy to prove that superposition (43) is characterized by
the following constant expectation value of energy:
\begin{equation}
<E> = (E_1+E_2)/2.
\end{equation}
On the other hand, the expectation value of passive gravitational
mass operator (11) for quantum state (43) is not constant and, as
possible to show, oscillates with time:
\begin{equation}
<\hat m^g_p> = m_e + \frac{E_1+E_2}{2 c^2} + \frac{V_{1,2}}{c^2}
\cos \biggl[ \frac{(E_1-E_2)t}{\hbar} \biggl],
\end{equation}
where we introduce matrix elements of the virial operator by the
following equation:
\begin{equation}
V_{1,2}= \int \Psi^*_1(r) \biggl(2 \frac{\hat {\bf
p}^2}{2m_e}-\frac{e^2}{r} \biggl) \Psi_2(r) d^3{\bf r} .
\end{equation}
Note that the mathematical origin of this result is that the
quantum virial theorem is valid for the expectation value of the
virial operator only in a stationary quantum state. From
Eqs.(45),(46), we can make a conclusion that the time-dependent
oscillations of the expectation value of passive gravitational
mass directly demonstrate breakdown of the equivalence between the
expectation values of passive gravitational mass and energy for
the superpositions of stationary quantum states. On the other
hand, it is easy to prove that the Einstein's equation is
fulfilled for averaged over time expectation values of passive
gravitational mass and energy. Indeed, from Eq.(45), it directly
follows that:
\begin{equation}
<<\hat m^g_p>>_t = m_e + \frac{E_1+E_2}{2 c^2} = \biggl<
\frac{E}{c^2} \biggl> .
\end{equation}

\subsection{5.2. Inequivalence between active gravitational mass and
energy}

Let us study the question if the expectation value of active
gravitational mass is equivalent to the expectation value of
energy for superpositions of stationary quantum states. For this
purpose, we again consider the simplest superposition of
stationary quantum states in a hydrogen atom, given by Eq.(43). As
we discussed in the previous subsection, the expectation value of
energy for such superposition (44) does not depend on time.
Nevertheless, it is easy to show that the expectation value of
electron active mass operator (39) oscillates with time:
\begin{equation}
<\hat m^g_a> = m_e + \frac{E_1+E_2}{2 c^2} + \frac{V_{1,2}}{c^2}
\cos \biggl[ \frac{(E_1-E_2)t}{\hbar} \biggl],
\end{equation}
where matrix element $V_{1,2}$ is given by Eq.(46). As seen from
Eq.(48), these time dependent oscillations are similar to that for
the expectation value of passive electron gravitational mass (45).
Eq.(48) directly demonstrates inequivalence between the
expectation values of active gravitational mass and energy for
superpositions of stationary quantum states. As we discussed
before such oscillations have a pure quantum mechanical origin and
do not have classical analogs. To simplify the situation, in the
same way as in the previous subsection, we can introduce the
averaged over time expectation value of active gravitational mass,
which obeys the Einstein's equation:
\begin{equation}
<<\hat m^g_a>>_t = m_e + \frac{E_1+E_2}{2 c^2} = \biggl<
\frac{E}{c^2} \biggl> .
\end{equation}

\section{6. First suggested idealized experiment}

Here, we suggest idealized experiment how to observe oscillations
of the expectation value of active gravitational mass of
electrons. By using laser, it is possible to create a macroscopic
ensemble of coherent superpositions of electron stationary states
in some gas [see Eq(43)]. They will be characterized by a feature
that each molecule has the same phase difference between two wave
function components, $\Psi_1(r)$ and $\Psi_2(r)$. In this case,
the macroscopic ensemble of atoms generates gravitational field,
which oscillates in time. In the case of hydrogen atoms, these
oscillations correspond to the following oscillations of the
active gravitational mass
\begin{equation}
M^g_a = n m_e + n \frac{E_1+E_2}{2 c^2} + n \frac{V_{1,2}}{c^2}
\cos \biggl[ \frac{(E_1-E_2)t}{\hbar} \biggl],
\end{equation}
where $n$ is the number of coherent atoms. The oscillating
gravitational field can be, in principle, measured by a probing
mass.

Let us discuss some characteristic features of oscillations (50),
including their relative magnitude. If we use actual numbers for a
hydrogen atom, we can obtain the following value for the matrix
element (46): $V_{1,2} \simeq 5.7 \ eV$. If we take into account
that $m_e c^2 \simeq 0.5 \ MeV$ and $m_p \simeq 1800 \ m_e$, we
can estimate relative strength of the oscillations (50): $\delta
m_e/m_e \sim 10^{-5}$ and $\delta m / m \sim 10^{-8}$. From these
estimations, we can conclude that the oscillations (50) are small
but not negligible. For the corresponding experiment it may be
important that they correspond to the following frequency: $\nu =
2.5 \times 10^{15}$ Hz. Here, we discuss what kind of molecules
have to be used for the above described experiment. As an
idealized example, we in the review always use a hydrogen atom or
a macroscopic ensemble of such atoms. In reality, as known, free
hydrogen atoms exist only in the Universe far from the Earth.
Therefore, more realistic examples are helium atoms or hydrogen
molecules. The corresponding calculations for a helium atom (see
Ref.[17]) show that all qualitative results of the review become
unchanged. Moreover, all estimations of the possible experimental
characteristics of oscillations (50), obtained in this section,
are valid. Nevertheless, there exists extra important for the
possible experiment point. Indeed, for creation of superpositions
of stationary quantum states by laser, we need that the
corresponding dipole matrix element will not be zero. On the other
hand, for the existence of the oscillations (50) we need non-zero
matrix element (46). Simple analysis shows that these are possible
only for molecules without center of symmetry [17]. We postpone
the detail discussions of the realistic experiments until the
corresponding calculations are complete. To summarize this
section, we hope that the oscillations of active gravitational
mass are experimentally found, despite the fact that they
correspond to quasi-stationary quantum states (43). If such
oscillations are experimentally observed this would be the first
direct observation of quantum effects in general relativity.

\section{7. Breakdown of the equivalence between passive gravitational
mass and energy for stationary quantum states}

In section 4, we have discussed that the expectation values of
both passive and active gravitational masses are equivalent to
energy for stationary quantum states. Nevertheless, as known,
quantum mechanics is not science about average values, but it is
science about an individual measurement. Therefore, in this
section, we review question about an individual measurement of
passive gravitational mass for a hydrogen atom. In particular, we
discuss in detail how Eqs.(10),(11) break the equivalence between
passive gravitational mass and energy at a microscopic level for
stationary quantum states [3-6]. First, let us pay attention that
passive gravitational mass operator for electron in a hydrogen
atom (11) does not commute with energy operator taken in the
absence of gravitational field. From physical point of view, this
means that quantum state of a hydrogen atom with definite energy
does not correspond to quantum state with definite passive
gravitational mass. In other words, measurements of gravitational
mass in such state can give different values which, as shown
below, are quantized [3-6]. In this section, we illustrate the
inequivalence between energy and passive gravitational mass by
means of two thought experiments.

\subsection{7.1. First thought experiment}
Here, we discuss the inequivalence between passive gravitational
mass and energy at a microscopic level for stationary quantum
states by considering the following thought experiment [4,5].
Suppose, there exists quantum state of a hydrogen atom with
definite energy in the absence of gravitational field (i.e, at $t
\rightarrow - \infty$),
\begin{equation}
\Psi_1(r,t) = \Psi_1(r) \exp(-im_ec^2t/\hbar -iE_1t/\hbar) \ .
\end{equation}
Then, we switch on adiabatically gravitational field (4) and,
therefore, in the presence of gravitational field (i.e., near
$t=0$), a hydrogen atom is characterized by the following wave
function:
\begin{equation}
\Psi(r,t) = \sum^{\infty}_{n = 1} a_n (t) \Psi_n(r)
\exp(-im_ec^2t/\hbar -i E_n t/\hbar) \ .
\end{equation}
[Note that, due to special symmetry of our problem, we need to
keep in Eq.(52) only normalized wave functions, $\Psi_n(r)$,
corresponding to isotropic $nS$ wave functions of the $n$-th
energy level in a hydrogen atom.]

Using Eqs.(10),(11), the adiabatically switched gravitational
field can be written as the following time-dependent perturbation
to quantum Hamiltonian in the absence of the field:
\begin{equation}
\hat U ({\bf r},R,t) = \phi(R)  \biggl[m_e + \biggl(\frac{\hat
{\bf p}^2}{2m_e}-\frac{e^2}{r}\biggl)/c^2 +\biggl(2 \frac{\hat
{\bf p}^2}{2m_e}-\frac{e^2}{r} \biggl)/ c^2 \biggl] \exp(\lambda
t) ,
\end{equation}
where $\lambda \rightarrow 0$. [It is important that our choice of
the perturbation in the adiabatical form (53) allows to avoid in
the Hamiltonian some extra terms related to velocities (see e.g.,
the Lagrangian in Ref.[9]).] Now, let us perform the standard
calculations by means of the quantum mechanical time dependent
perturbation theory. As a result, we obtain for amplitudes in
Eq.(52) the following expressions:
\begin{equation}
a_1(t)=\exp \Big[ - \frac{i \phi(R) m_e c^2 t + i \phi(R) E_1  t
}{c^2 \hbar} \Big] \  ,
\end{equation}

\begin{equation}
a_n(0)=  - \frac{\phi(R)}{c^2} \frac{V_{1,n}}{E_n-E_1} \  , \ n
\neq 1 \ ,
\end{equation}
where
\begin{equation}
V_{1,n}= \int \Psi^*_1(r) \biggl( 2 \frac{{\bf \hat p}^2}{2 m_e} -
\frac{e^2}{r} \biggl) \Psi_n(r) d^3 {\bf r} \ .
\end{equation}
[We pay attention to the fact that perturbation (53) has the
following selection rule: electrons from 1S ground state of a
hydrogen atom can be excited only into nS excited level.] Here, we
discuss in detail Eqs.(54)-(56). We note that Eq.(54) represents
the well known red shift of atomic ground state energy level in
gravitational field. It is important that amplitudes (55)
correspond to some novel effect. Indeed, they show that there is
non-zero probability,
\begin{equation}
P_n = |a_n(0)|^2 = \Big[ \frac{\phi(R)}{c^2} \Big]^2 \ \Big(
\frac{V_{1,n}}{E_n-E_1} \Big)^2 \ , \ n \neq 1,
\end{equation}
that, in the presence of gravitational field (i.e., at $t=0$),
electron occupies $nS$ energy level in a hydrogen atom. In other
words, this demonstrates that quantum measurements of passive
electron gravitational
 mass (11) in quantum state with definite energy (52) can result in the
 following quantized values:
 \begin{equation}
m^g_p (n) = m_e + E_n / c^2 \ ,
\end{equation}
which correspond to probabilities (57). Note that probabilities
(57) are of the second order with respect to the small parameter,
$|\phi(R)|/c^2 \ll 1$, and, therefore, they are rather small.
Nevertheless, we pay attention that the corresponding changes of
electron passive gravitational mass are comparably large. Indeed,
they do not contain small parameter, $|\phi(R)|/c^2 \ll 1$, and
are of the order of $\delta m^g_p / m^g_p \sim \alpha^2 \sim
10^{-4}$, where $\alpha$ is the fine structure constant. It is
important that quantization law (58) can be measured indirectly.
The point is that the all excited atomic levels are
quasi-stationary and, thus, the quantization law (58) can be
discovered by measuring electromagnetic radiation from macroscopic
ensemble of the atoms. This procedure is discussed in more detail
in section (8).

\subsection{7.2. Second thought experiment}

To further study phenomenon of the inequivalence between passive
gravitational mass and energy for stationary quantum states at the
microscopic level, let us consider the second possible thought
experiment [3,5]. Suppose that, at $t=0$, we create, in
gravitational field (4), electron stationary quantum state,
corresponding to ground state of a hydrogen atom in the absence of
gravitational field [see Eq.(51)]. Then, we take into account
gravitational field and find that the wave function (51) is not
anymore a ground state of the Hamiltonian (10),(11). It is
important that for proper local observer, in accordance with (5)
and main principles of general relativity, a general solution of
the Schr\"{o}dinger equation in the presence of the gravitational
field (4) is
 \begin{equation}
\Psi(r,t) = (1-\phi/c^2)^{3/2} \sum^{\infty}_{n = 1} a_n \Psi_n
[(1- \phi/c^2)r] \times \exp[-i m_e c^2 (1+\phi/c^2) t/\hbar]
\exp[-i E_n(1+\phi/c^2) t/\hbar] \ .
\end{equation}
It is easy to show that wave function (59) is a series of
eigenfunctions of passive gravitational mass operator (11) if we
restrict ourself only linear terms with respect to the parameter
$\phi/c^2$. We note that, in Eq.(59), factor $(1-\phi/c^2)$ in a
hydrogen atom wave functions, $\Psi_n[(1-\phi/c^2)r]$, is due to a
curvature of space and the total wave function (59) is normalized.
Another factor $(1+\phi/c^2)$ corresponds to the famous red shift
of electron energy levels in the gravitational field (4).

According to the main principles of quantum mechanics, probability
that electrons with wave function (51) occupies excited quantum
state with energy $m_e c^2 (1+\phi/c^2) + E_n(1+\phi/c^2)$ can be
written as
\begin{equation}
P_n = |a_n|^2, \ a_n = \int \Psi^*_1(r) \Psi_n [(1-\phi/c^2)r] d^3
{\bf r} \approx - ( \phi/c^2) \int \Psi^*_1(r) r \Psi'_n(r) d^3
{\bf r}.
\end{equation}
[It is important that the calculations show that for $a_1$ in
Eq.(60) the linear term with respect to gravitational potential
$\phi$ is zero. This corresponds to the statement of section 4
about the equivalence of the expectation values of passive
gravitational mass and energy and is a consequence of the quantum
virial theorem.] If we take into account the fact that the
Hamiltonian is a Hermitian operator, we can show that, for $n \neq
1$,
\begin{equation}
\int \Psi^*_1(r) r \Psi'_n(r) d^3 {\bf r} = V_{1,n}/ (E_n - E_1) ,
\end{equation}
where the matrix element of the virial operator is given by
Eq.(56). Here, we discuss physical meaning of Eqs.(60),(61). We
point out that they directly demonstrate that there exist a
non-zero probability,
\begin{equation}
P_n = |a_n|^2 = \Big[ \frac{\phi(R)}{c^2} \Big]^2 \ \Big(
\frac{V_{n,1}}{E_n-E_1} \Big)^2 \ , \ n \neq 1,
\end{equation}
that, in the presence of gravitational field, electron occupies nS
atomic orbital, which directly breaks the expected Einstein's
equation, $m^g_p=m_e+E_1/c^2$. Let us discuss this question in
more detail. Indeed, non-zero probabilities (62) demonstrate that
there are situations, where quantum measurements of electron
passive gravitational mass in state with definite energy give the
quantized values (58). Note that, as it has to be, the
probabilities (57) and (62) are equal, which show the equivalence
of the first and second thought experiments. We pay attention that
the probabilities (57),(62) are quadratic with respect to small
parameter, $|\phi(R)|/c^2 \ll 1$. It is easy to make sure that, to
calculate the probabilities (62) with such accuracy for $n \neq
1$, we need to calculate wave functions (52),(54),(55) in linear
approximation with respect to the above mentioned small parameter.
Moreover, analysis shows that calculation of wave function (59)
with accuracy of $|\phi(R)|^2/c^4$ will change electron passive
gravitational mass of the order of $\delta m^g_p \sim
(\phi/c^2)m_e \sim 10^{-9} m_e$. This value is much smaller than
the characteristic difference between the quantized values in
Eq.(58), $\delta m^g_p \sim \alpha^2 m_e \sim 10^{-4} m_e$. We
also would like to stress that the existence of the small
probabilities (57),(62), $P_n \sim 10^{-18}$, to break the
Einstein's equation for passive gravitational mass and energy do
not contradict to the existing E\"{o}tvos type experimental
measurements [11], confirming the equivalence principle with
accuracy of the order of $\delta m/m \sim 10^{-13}$. As mentioned
in the previous subsection, an important role plays the fact that
all excited levels of a hydrogen atom are not strictly stationary.
Therefore, quantization law (58) can be measured by optical
methods, which are much more sensitive that the E\"{o}tvos type
experiments.

\section{8. Second suggested idealized experiment}
The second suggested idealized experiment [3-5] is designed to
observe the passive gravitation mass quantization law (58) by
measuring radiation, emitted by a macroscopic ensemble of some
atoms supported and moved in the Earth's gravitational field. All
calculations below are done for hydrogen atoms, although in
practice helium atoms or hydrogen molecules can be used [17]. We
cannot except that some kind of solids may be more convenient to
use in such kind of experiments, but this discussion has to be
postponed until more detailed calculations are done. As mentioned
before, in this review, we restrict ourself by the following
idealized experimental conditions. There is a small spacecraft,
which carries a tank with hydrogen atoms. The atoms do not
interact with each other and are supported in the Earth's
gravitational field and moved with constant velocity in the field.

\subsection{8.1. Lagrangian and Hamiltonian}
Here, we review in detail derivation of the Hamiltonian [3],
corresponding to the suggested idealized experiment [3-6]. To this
end, we first consider the Lagrangian for the following three
bodies: electron and proton, bound by the Coulomb electrostatic
field, and the Earth moving from the atom with small constant
velocity, $u \ll c$. This case for classical model of a hydrogen
atom is considered in Ref.[9] and, therefore, we can use the
result [9] that the above mentioned three-body Lagrangian can be
written as sum
\begin{equation}
L = L_{kin} + L_{e} + L_{G} + L_{e,G},
\end{equation}
where $L_{kin}$, $L_{e}$, $L_{G}$, and $L_{e,G}$ are kinetic,
electric, gravitational, and electro-gravitational parts of the
Lagrangian, correspondingly. We recall that, in our approximation,
we omit in the Hamiltonian and Lagrangian all terms of the order
of $\phi^2/c^4$ and $(v/c)^4$. In other words, we keep only
classical kinetic energy and the electrostatic Coulomb energy
interactions with external gravitational field. In this case four
contributions to the Lagrangian (63) can be written in the
following simplified forms:
\begin{equation}
L_{kin} + L_{e} = - Mc^2 - m_pc^2 - m_ec^2 + M\frac{u^2}{2} +
m_e\frac{{\bf v}^2}{2} + \frac{e^2}{r} \ ,
\end{equation}

\begin{equation}
L_{G} = G \frac{m_p M}{(R+ut)} + G \frac{m_e M}{(R+ut)} \biggl\{1
- \frac{1}{2} [{\bf v} \cdot {\bf u} + ({\bf v} \cdot \hat{\bf
R})({\bf u} \cdot \hat{\bf R})]/c^2 \biggl\} + \frac{3}{2} G
\frac{m_e M}{(R+ut)} \frac{{\bf v^2}}{c^2} \ ,
\end{equation}

\begin{equation}
L_{e,G} = - 2 G \frac{M}{(R+ut)c^2} \frac{e^2}{r} \ ,
\end{equation}
where $\hat{\bf R}$ is a unit vector directed along ${\bf R}$ and
where we use the inequality $m_e \ll m_p$.

For further development, it is natural to suppose that the
spacecraft
 velocity, $u$, is small enough comparable to the typical electron
 velocity in atoms and molecules, $v \sim \alpha c$:
\begin{equation}
u \ll v \sim \alpha c \ .
\end{equation}
Using Eq.(67), we simplify the Lagrangian (63)-(66) in the
following way:
\begin{equation}
L = m_e \frac{v^2}{2} + \frac{e^2}{r} - \frac{\phi(R+ut)}{c^2}
\biggl[ m_e + 3 m_e \frac{{\bf v}^2}{2} - 2 \frac{e^2}{r} \biggl]
\ .
\end{equation}
[We pay attention that, in the Lagrangian (68), we disregard the
fact that the center of mass of a hydrogen is not exactly
coincides with proton. It is possible to show that, doing so, we
disregard terms of the order of $|\phi(R+ut)|m_evu/c^2 \ll
|\phi(R+ut)|m_ev^2/c^2$. We also use electron mass, instead of the
so-called reduced mass, which is possible for $m_e \ll m_p$.] From
Eq.(68), it is easy to find the corresponding Hamiltonian:
\begin{equation}
H= \frac{{\bf p}^2}{2m_e} -\frac{e^2}{r} + \frac{\phi(R+ut)}{c^2}
\biggl[ m_e + 3 \frac{{\bf p}^2}{2m_e} - 2 \frac{e^2}{r} \biggl] \
,
\end{equation}
which can be easily quantized:
\begin{equation}
\hat H= \frac{\hat {\bf p}^2}{2m_e} -\frac{e^2}{r} +
\frac{\phi(R+ut)}{c^2} \biggl[ m_e + 3 \frac{\hat {\bf p}^2}{2m_e}
- 2 \frac{e^2}{r} \biggl] \ .
\end{equation}

\subsection{8.2. Mass quantization law and emitted photons}

Once the quantum Hamiltonian (70) is known, we can consider theory
of the second suggested idealized experiment. We assume that at
initial moment of time, $t=0$, all hydrogen atoms are in a ground
state and located in small spacecraft, which is placed at distance
$R'$ from the Earth's center. The wave function of each hydrogen
atom in this case can be written in the following form [see
Eq.(59)]:
\begin{equation}
\tilde{\Psi}_1(r,t) = (1-\phi'/c^2)^{3/2} \Psi_1[(1-\phi'/c^2)r]
\times \exp[-im_ec^2(1+\phi'/c^2) t /\hbar] \
\exp[-iE_1(1+\phi'/c^2)t/\hbar] \ ,
\end{equation}
where $\phi'=\phi(R')$. Since we consider non-interacting hydrogen
atoms, here we discuss optical properties of a single atom and
later will take into account the fact that we have a macroscopic
ensemble to the atoms. We recall that the atom is supported in
gravitational filed and moved from the Earth with constant
velocity, $u \ll \alpha c$ [see Eq.(67)], by spacecraft. From
Eq.(70), it follows that electron wave function at latter moments
of time in the inertial coordinate system, related to the
spacecraft, can be expressed as
 \begin{equation}
\tilde{\Psi}(r,t) = (1- \phi'/c^2)^{3/2} \sum^{\infty}_{n=1}
\tilde{a}_n(t) \Psi_n[(1-\phi'/c^2)r] \exp[-im_ec^2(1+\phi'/c^2) t
/\hbar]
 \ \exp[-iE_n(1+\phi'/c^2)t/\hbar] ,
\end{equation}
where the perturbation for "free" Hamiltonian of a hydrogen atom
is
\begin{equation}
\hat U ({\bf r},t) =\frac{\phi(R'+ut)-\phi(R')}{c^2}  \biggl(3
\frac{\hat {\bf p}^2}{2m_e}-2\frac{e^2}{r} \biggl) .
\end{equation}
(It is important to stress that in the spacecraft, moving with
constant velocity, each hydrogen atom and electron experience
gravitational force. In this section, we suggest the experiment,
where this force is compensated by some forces of
non-gravitational origin. It is possible to show that this causes
small changes of the atomic energy levels, which are not important
for the further calculations. Since the gravitational forces are
compensated, the atom does not feel acceleration, ${\bf g}$, but
rather feels the time dependent gravitational potential,
$\phi(R'+ut)$ [see Eq.(73)]).

Now, let us apply the standard time-dependent quantum mechanical
perturbation theory to determine amplitudes $\tilde a_n(t)$ in
Eq.(72):
\begin{equation}
\tilde{a}_n(t)= -\frac{V_{1,n}}{\hbar \omega_{n,1}c^2}
\biggl\{[\phi(R'+ut)-\phi(R')]  \exp(i \omega_{n,1}t) - \frac{u}{i
\omega_{n,1}} \int^t_0 \frac{d \phi(R'+ut)}{dR'} d[\exp(i
\omega_{n,1}t)]\biggl\}, \ n \neq 1 \ ,
\end{equation}
where
\begin{equation}
\omega_{n,1} = (E_n - E_1)/\hbar
\end{equation}
and $V_{1,n}$ is given by Eq.(56). We point out that, in the
suggested experiment, the following condition is valid:
\begin{equation}
u \ll \omega_{n,1} R \sim \alpha c (R/r_B) \sim 10^{13} c.
\end{equation}
It is easy to prove that Eq.(76) means that we can omit the second
term in Eq.(74):
\begin{equation}
\tilde{a}_n(t)= -\frac{V_{1,n}}{\hbar \omega_{n,1}c^2}
[\phi(R'+ut)-\phi(R')]  \exp(i \omega_{n,1}t) \ , \ n \neq 1 .
\end{equation}
Let us discuss the main consequences of Eq.(77). As directly seen
from (77), if the excited levels of the atom were strictly
stationary, then there would exist possibilities to find the
quantized values of passive gravitational mass with $n \neq 1$ in
Eq.(58). They correspond to the following probabilities:
\begin{equation}
\tilde{P}_n(t)= \biggl( \frac{V_{1,n}}{\hbar \omega_{n,1}}
\biggl)^2 \frac{[\phi(R'+ut)-\phi(R')]^2}{c^4} = \biggl(
\frac{V_{1,n}}{\hbar \omega_{n,1}} \biggl)^2
 \frac{[\phi(R")-\phi(R')]^2}{c^4} \ , n \neq 1,
\end{equation}
where $R"=R'+ut$. In reality, the excited levels of a hydrogen
atom are quasi-stationary, therefore, they spontaneously decay
with time. In this case, it is possible to measure the breakdown
of the Einstein's equation indirectly by measuring electromagnetic
radiation, emitted by a macroscopic ensemble of the atoms. Then,
Eq.(78) corresponds to probability that a hydrogen atom emits a
photon with frequency (75) during the time interval $t$. [We would
like to stress here that the dipole matrix element for quantum
transition $nS \rightarrow 1S$ is zero. In this situation, we can
expect quadrupole transition $2S \rightarrow 1S$ and dipole
transitions like $3S \rightarrow 2P$, ect.]

Important fact is that the probabilities (78) can be written in
the form, where they depend only on gravitational potentials at
the initial and final points of spacecraft trajectories, $\phi' =
\phi (R')$ and $\phi" = \phi (R")$. This allow to determine their
physical meaning. Below, we consider a general solution of the
Hamiltonian (10),(11) at final position of spacecraft, where it is
located at distance $R"$ from center of the Earth:
\begin{equation}
\tilde{\Psi}(r,t) = (1- \phi"/c^2)^{3/2} \sum^{\infty}_{n=1}
\tilde{a}_n \Psi_n[(1-\phi"/c^2)r] \exp[-im_ec^2(1+\phi"/c^2) t
/\hbar]
 \ \exp[-iE_n(1+\phi"/c^2)t/\hbar] .
\end{equation}
We pay attention that wave function (79) is an infinite series of
the eigenfunctions corresponding to a definite weight (i.e.,
passive gravitational mass) at distance $R"$ from the Earth's
center. If we take into account that the wavefunction (71)
corresponds to quantum state with definite energy at distance $R'$
from the Earth's center, then, according to the main principles of
quantum mechanics, the expression,
\begin{eqnarray}
&a_n = \int \Psi^*_1[(1-\phi'/c^2)r] \Psi_n [(1-\phi"/c^2)r] d^3
{\bf r} \approx - [(\phi"-\phi')/c^2] \int \Psi^*_1(r) r
\Psi'_n(r) d^3 {\bf r}\ ,
\nonumber\\
&P_n = |a_n|^2 = \frac{(\phi"-\phi')^2}{c^4}\biggl|\Psi^*_1(r) r
\Psi'_n(r) d^3 \biggl|^2 \ ,
\end{eqnarray}
represents the probability that electron with definite energy at
position $R'$ has the quantized passive gravitational mass (58)
values with $n \neq 1$. Now, let us use Eq.(61) and rewrite
probabilities (80) in a familiar way,
\begin{equation}
P_n= \biggl( \frac{V_{n,1}}{\hbar \omega_{n,1}} \biggl)^2
 \frac{[\phi(R")-\phi(R')]^2}{c^4} \ , n \neq 1 ,
\end{equation}
coinciding with Eq.(78). Here, we can make a conclusion that all
photons, emitted during spacecraft motion between positions $R'$
and $R"$, are due to breakdown of the Einstein's equation for
passive gravitational mass and energy (i.e., they are due to
quantization of gravitational mass [see Eq.(58)]). Below, we
estimate the probabilities (78). It is easy to see that, by using
spacecraft, we can reach the condition $|\phi(R")| \ll
|\phi(R')|$. In this case Eq.(78) coincides with Eqs.(57),(62) and
can be written as
\begin{equation}
\tilde{P}_n = \biggl( \frac{V_{1,n}}{E_n - E_1} \biggl)^2
\frac{\phi^2(R')}{c^4}  \simeq  0.49 \times 10^{-18} \biggl(
\frac{V_{n,1}}{E_n-E_1} \biggl)^2 ,
\end{equation}
where we make use of the following values of the Earth's radius
and mass, $R_0 \simeq 6.36 \times 10^6 m$ and $M \simeq 6 \times
10^{24} kg$, respectively. Characteristic feature of probability
(82) is that it is quadratic with respect to small parameter,
$|\phi (R')| \ll 1$, and, thus, small, $P_n \sim 10^{-18}$.
Nevertheless, the number of the emitted photons can be very large
since $V_{n,1}/(E_n-E_1) \sim 1$ and the Avogadro number (i.e., in
our case, the number of atoms in one mole of atomic hydrogen) is
$N_A = 6 \times 10^{23}$. The calculations show that, for 1000
moles of the atoms, the number of emitted photons is estimated as
\begin{equation}
N_{n,1} = 2.95 \times 10^{8} \biggl( \frac{V_{n,1}}{E_n-E_1}
\biggl)^2 , \ N_{2,1} = 0.9 \times 10^8 .
\end{equation}
Presumably such large amount of photons can be experimentally
detected, where $N_{n,1}$ determines the number of photons,
emitted with frequency (75).

\section{9. Summary}
In the review, we have discussed in detail the question about
active and passive gravitational masses of a composite quantum
body [3-6], using a hydrogen atom in the Earth's gravitational
field as the simplest example. In particular, we have shown that
the expectation values of active and passive gravitational masses
are equivalent to the expectation value of energy for stationary
quantum states. On the other hand, this equivalence has been
discussed to be broken for non-stationary quantum states, where
the expectation values of energy are constant, but time-dependent
oscillations of the expectation values of the masses appears. We
have suggested idealized experiment how to detect these
time-dependent oscillations of active gravitational mass. Here, we
pay attention to the following difficulty to be resolved before
such experiment is done. The point is that the non-stationary
quantum states decay with time. Therefore, it is necessary to
distinguish experimentally weak oscillating gravitational signal
from strong electromagnetic radiation effects. For stationary
quantum states, we have also discussed breakdown of the
equivalence between passive gravitational mass and energy at a
microscopic level. We have illustrated this phenomenon by
discussing two thought experiments and one idealized. Our
idealized experiment also needs an improvements and some
additional calculations. First, the existing calculations [3-6]
have to be extended to the cases of helium and molecular hydrogen,
since large amounts of atomic hydrogen exist only in the Universe
very far from the Earth. Second, effects of atomic (molecular)
interactions have to be taken into account. We cannot even exclude
that it maybe more experimentally effective to use some solid
state bodies, instead of a tank of a gas, suggested above and in
Refs.[3-6]. We hope that all the above mentioned and possible
other experimental difficulties are solved and the discussed above
two experiments are done. They would be the first direct
experimental observations of quantum effects in general
relativity.

\section*{Acknowledgments}

We are thankful to N.N. Bagmet (Lebed), V.A. Belinski, Steven
Carlip, Fulvio Melia, Douglas Singleton, and V.E. Zakharov for
fruitful and useful discussions. This work was supported by the
NSF under Grant DMR-1104512.

$^*$Also at: L.D. Landau Institute for Theoretical Physics, RAS,
2 Kosygina Street, Moscow 117334, Russia.



\vspace{6pt}



\begin{thebibliography}{0}    

\bibitem{Collela} R. Collela, A.W. Overhauser, and S. Werner, {\it Phys.
Rev. Lett.} {\bf 34} (1975) 1472.

\bibitem{Nesvizhevsky} V.V. Nesvizhevsky, H.G. Borner, A.K. Petukhov et
al., {\it Nature} {\bf 415} (2002) 297.

\bibitem{Lebed-1} A.G. Lebed, {\it Cent. Eur. J. Phys.} {\bf 11} (2013)
969.

\bibitem{Lebed-2} A.G. Lebed, {\it J. Phys.: Conf. Ser.} {\bf 490} (2014)
012154.

\bibitem{Lebed-3} A.G. Lebed, {\it Adv. High Ener. Phys.} {\bf 2014}
(2014) 678087.

\bibitem{Lebed-4} A.G. Lebed, Breakdown of the equivalence between
passive gravitational mass and energy for a quantum body, in {\it
Proc. of the 13th Marcel Grossmann Meeting on General Relativity},
eds. Remo Ruffini, Robert Jantzen, and Kjell Rosquist (World
Scientific, Singapore, 2014), p.~1953.



\bibitem{Landau} See, for example, L.D. Landau and E.M. Lifshitz,
{\it The Classical Theory of Fields}, 4th edn.
(Butterworth-Heineman, Amsterdam, 2003).

\bibitem{Misner} C.W. Misner and P. Putnam, {\it Phys. Rev.} {\bf 116}
(1959) 1045.

\bibitem{Nordtvedt} K. Nordtvedt, {\it Class. Quan. Grav.} {\bf 11}
(1994) A119.


\bibitem{Carlip} S. Carlip, {\it Amm. J. Phys.} {\bf 66}
(1998) 409.

\bibitem{Wheeler} See, for example, C.W. Misner, K.S. Thorne, and J.A.
Wheeler, {\it Gravitation}, (W.H. Freeman and Co, San Francisco,
1973).


\bibitem{Fishbach} E. Fischbach, B.S. Freeman, W.K. Cheng,
{\it Phys. Rev. D} {\bf 23} (1981) 2157.

\bibitem{Virial-1}
See, for example, D. Park, {\it Introduction to Quantum Theory},
3rd edn. (Dover, New York, 2005).

\bibitem{Relativistic}
See, for example, E. Schwable, {\it Advanced Quantum Mechanics},
3rd edn. (Springer, Berlin, 2005).

\bibitem{Virial-2} W. Lucha and F.F. Schoberl, {\it Phys. Rev. Lett.}
{\bf 64} (1990) 2733.

\bibitem{Semiclassical-1}
See, for example, N.D. Birrell and P.C.W. Devis, {\it Quantum
Fields in Curved Space}, 3rd edn. (Cambridge University Press,
Cambridge, UK, 1982).

\bibitem{Lebed-5}
A.G. Lebed, in a preparation.




\end{thebibliography}
\end{document}